\documentclass{mn2e}
\usepackage{graphicx}

\newcommand{\tmin}{{T_{\rm min}}}
\newcommand{\msun}{{\rm M}_{\odot}}
\newcommand\ion[2]{#1$\;${\small\rmfamily{#2}}\relax}

\title[Star Formation in Ram Pressure Stripped Tails]{Star Formation in Ram Pressure Stripped Galactic Tails}
\author[S. Tonnesen and G. L. Bryan]{Stephanie Tonnesen$^{1}$\thanks{E-mail:
stonnes@astro.princeton.edu (ST)); gbryan@astro.columbia.edu (GLB)} and Greg L. Bryan$^{2}$\\
$^{1}$Department of Astrophysics, Princeton University, Peyton Hall, Princeton, NJ 08544, USA\\
$^{2}$Department of Astronomy, Columbia University, Pupin Physics Laboratories, New York, NY 10027, USA}

\begin{document}

\pagerange{\pageref{firstpage}--\pageref{lastpage}} \pubyear{2011}

\maketitle

\label{firstpage}

\begin{abstract}
We investigate the impact of star formation and feedback on ram pressure stripping using high-resolution adaptive mesh simulations, building on a previous series of papers that systematically investigated stripping using a realistic model for the interstellar medium, but without star formation.  We find that star formation does not significantly affect the rate at which stripping occurs, and only has a slight impact on the density and temperature distribution of the stripped gas, indicating that our previous (gas-only) results are unaffected.  For our chosen (moderate) ram pressure strength, stripping acts to truncate star formation in the disk over a few hundred million years, and does not lead to a burst of star formation.  Star formation in the bulge is slightly enhanced, but the resulting change in the bulge-to-disk ratio is insignificant.  We find that stars do form in the tail, primarily from gas that is ablated from the disk and the cools and condenses in the turbulent wake.  The star formation rate in the tail is low, and any contribution to the intracluster light is likely to be very small.  We argue that star formation in the tail depends primarily on the pressure in the intracluster medium, rather than the ram pressure strength.  Finally, we compare to observations of star formation in stripped tails, finding that many of the discrepancies between our simulation and observed wakes can be accounted for by different intracluster medium pressures.
\end{abstract}



\section{Introduction}

As galaxies orbit within a cluster, their interstellar medium (ISM) may interact directly with the intracluster medium (ICM), the hot halo of gas bound by the cluster gravitational potential.  This is often thought of as a gas-only affair, in which stars remain unaffected. For example, ram pressure stripping (and related processes) by the ICM is only able to remove a galaxy's gas (Gunn \& Gott 1972).  Although stars are not directly affected, there is increasing evidence that ISM-ICM interactions do impact star formation.

In general, galaxies in clusters have lower star formation rates than galaxies of the same morphological type in the field (e.g. Hashimoto et al. 1998; Rines et al. 2005; Balogh et al. 1998; although Gom\'ez et al. (2003) found that a galaxy's star formation rate depended more on local galaxy density than on cluster membership).  Gavazzi et al. (2006) found that the star formation rate in cluster galaxies was related to the amount of \ion{H}{I}:  galaxies with normal \ion{H}{I} had twice the H$\alpha$ equivalent widths of \ion{H}{I} deficient galaxies.  Koopmann \& Kenney (2004) found that Virgo spirals are forming stars primarily in the centres of their disks, which can be explained by the outer gas disk having been stripped by the ICM.  In a more recent study of ten Virgo spiral galaxies, Crowl \& Kenney (2008) find that the star formation history and quenching time of five of their galaxies is consistent with ram pressure stripping in the cluster centre, while the other five have more complicated histories.

Although there is evidence linking ISM-ICM interactions to star-formation quenching, the question of whether interactions with the ICM could also induce star formation in a galactic disk remains open.  For example, star formation could be triggered by the increase in surrounding pressure when a galaxy enters a high-density ICM (Dressler \& Gunn 1983; Evrard 1991; Fujita 1998; Smith et al. 2010).  Fujita \& Nagashima (1999) found ram pressure induced star formation in simulated galaxies.  

Observations of post-starburst galaxies in clusters indicate that star formation can be induced by environmental processes (Dressler \& Gunn 1983), although the exact mechanism is still unknown.  Post-starburst galaxies reside preferentially in clusters at $z = 0.3-0.6$ (e.g. Poggianti et al. 1999, Poggianti et al. 2004; Tran et al. 2004; but see Balogh \& Bower 2003).  Although most observations at both lower and higher redshift have found k+a fractions increasing with decreasing galaxy density (e.g. Zabludoff et al. 1996; Hogg et al. 2006; Goto 2007; Yang et al. 2008), Poggianti et al. (2004) find a large population of low-luminosity post-starburst galaxies in Coma.  Tran et al. (2003) examined E+A galaxies in three intermediate-redshift clusters and found that the majority were not associated with mergers.  By considering the spatial distribution of post-starburst galaxies in clusters, some observers have found evidence that both the star formation quenching and earlier starburst could be related to interactions with the ICM (Poggianti et al. 2004; Poggianti et al. 2009; Ma \& Ebeling 2008).   

By investigating whether and how ram pressure stripping by a dense ICM can affect star formation in a galaxy, we can shed light on what drives the evolution of cluster galaxies.  From z$\sim$0.5 to z=0, much of the morphological evolution in clusters has been from spirals to S0s (Dressler et al. 1997).  In order for a spiral galaxy to evolve into an S0, both spectroscopic and morphological changes must take place:  the galaxy must become red, it must lose spiral structure in its disk, and the bulge-to-disk ratio must increase.  

Ram pressure stripping, through gas removal, can cause a galaxy to become red.  Passive spirals, red spirals without star formation, have been observed to reside preferentially in clusters (Moran et al. 2007; Poggianti et al. 1999).  A galaxy-ICM interaction is the likely mechanism for forming passive spirals because these galaxies have had their gas removed but are morphologically spirals (because their stellar disks are relatively undisturbed).  If passive spirals are precursors to S0s, then a galaxy-ICM interaction is a step in the evolution of normal spirals into S0s.  

Ram pressure stripping can also result in a disk galaxy losing its spiral arms.  Bekki et al. (2002) used simulations to show that if gas is no longer accreted by a galaxy, it will lose its spiral arms in about 3.5 Gyr.  This is in good agreement with observational and model estimates for how long morphological transformation may take (e.g Kodama \& Smail 2001; Poggianti et al. 1999; but see Moran et al. 2007 for a shorter estimate). 

Finally, a galaxy's bulge-to-disk (B/D) ratio can increase either by fading the disk or growing the bulge.  Ram pressure stripping can result in the disk of a galaxy fading.  Fading the disk of a galaxy with a B/D = 0.2 will result in a galaxy with a B/D = 0.5, so an Sb galaxy, after disk fading, will have the B/D ratio of an S0 galaxy (Fujita \& Nagashima 1999; Solanes et al. 1989).  Solanes et al. (1989) find that the bulge luminosities of Sa galaxies are similar to those of S0s, and all galaxy types have a range of luminosities, so it is not universally necessary to increase the bulge luminosity to transform a spiral to an S0.  However, if the slow fading of the disk was the only mechanism at work, the total luminosity of S0s should be less than that of spirals, which is not generally the case (Burstein et al. 2005).  Christlein \& Zabludoff (2004) compared the bulge and disk luminosities of cluster galaxies, and concluded that S0s in clusters form by growing the bulge of spiral galaxies.  Therefore it is important to know whether ram pressure can induce star formation that may grow a galaxy's bulge.

There can also be star formation in a ram pressure stripped tail of gas, even though molecular clouds are considered too dense to be directly stripped from a galactic disk.  It is still unclear how common star formation in stripped gas tails is, as many tails observed in \ion{H}{I} do not have any associated star formation.  For example, several one-sided \ion{H}{I} tails have been observed in Virgo.  Chung et al. (2007) found 7 one-sided tails between 0.6-1 Mpc in projected distance from M87.  Oosterloo \& van Gorkom (2005) found a $\sim$110 kpc \ion{H}{I} tail associated with NGC 4388.  NGC 4438 was originally believed to be an interacting galaxy (e.g. Hibbard \& van Gorkom 1990; Kenney et al. 1995; Vollmer et al. 2005), but more recent work has concluded that it is likely only ram pressure stripped.  NGC 4522 is 1 Mpc from M87, and Kenney et al. (2004) conclude that it is being ram pressure stripped by an overdense or moving region of the ICM.  Finally, NGC 4402 also has an \ion{H}{I} tail (Crowl et al. 2005).  Of these eleven galaxies with clear stripping signatures in \ion{H}{I} observations, only four have been found to have star formation either from UV emission or \ion{H}{II} regions.

The four \ion{H}{I} tails with star formation tend to have short stellar tails with low star formation rates.  Abramson et al. (2011) find 9 UV emitting regions near NGC 4330, for a total of 4.55 $\times$ 10$^6$ M$_\odot$ of extragalactic stars.  Cortese et al. (2003; 2004) found an \ion{H}{II} region 3 kpc above the disk of NGC 4402 in the Virgo cluster.  Similarly, NGC 4522, and NGC 4438 have ongoing star formation in their stripped tails close to the galaxy (Kenney \& Koopmann 1999; Boselli et al. 2005).  NGC 4438 has the most distant UV emission out to nearly 30 kpc from the disk (Boselli et al. 2005).  Although Gerhard et al. (2002) find an \ion{H}{II} region near NGC 4388 in the Virgo cluster, this \ion{H}{II} region is not near the long \ion{H}{I} tail observed by Oosterloo \& van Gorkom (2005), and there have been no stars found associated with the stripped \ion{H}{I}.

There have been an increasing number of recent observations of star formation in stripped gas tails.  Long trails of star-forming knots were observed in two massive galaxy clusters by Cortese et al. (2007), extending as far as 80 kpc from one of the galaxies.  Cortese et al. (2007) find that a combination of tidal and ram pressure stripping are affecting the galaxies.  In the Coma cluster, Yoshida et al. (2008) found a complex of  H$\alpha$ filaments and clouds extending up to 80 kpc from the E+A galaxy RB 199.  They also conclude that the most likely gas removal scenario involves a combination of a merger and ram pressure stripping.  Yagi et al. (2010) find another 13 galaxies with young stars or H$\alpha$ clouds in tails.  Finally, ES0 137-001, which we discussed in detail in Tonnesen et al. (2011), has 35 HII regions extending more than 30 kpc from the disk.  These HII regions are spatially correlated with a ram pressure stripped tail of gas (Sun et al. 2006, 2007, 2010).  Hester et al. (2010) found H$\alpha$ emission indicative of star formation in the UV tail of IC 3418 (Martin et al. 2005), a low surface brightness galaxy in the Virgo Cluster.

In a series of 12 SPH simulations including ram pressure stripping and star formation, Kapferer et al. (2009) found that stars can form in their ram pressure-stripped tail out to hundreds of kiloparsecs from the disk.  In fact, using fast, high-density ICM winds, the authors found more stars forming in the stripped tail than in the remaining disk.  The star formation rate increased due to enhanced external pressure provided by the ICM.  They also found that stars formed in the tail could fall back into the stellar bulge.  

Clearly, the standard lore that galaxy-ICM interactions do not result in stars outside of the galaxy has been overturned.  In fact, if star formation in stripped tails is common, ram pressure stripping could contribute a significant fraction of the Intracluster Light (ICL).

In this paper, we run a set of high resolution simulations (about 40 pc resolution, which is small enough to marginally resolve giant molecular clouds) to understand whether ram pressure can induce star formation in a galactic disk, producing starburst galaxies or increasing the mass of the bulge, and whether stars can form in a stripped gas tail and add to the ICL.  

The paper is structured as follows.  After a brief introduction to our methodology, we provide the general characteristics of our simulation (\S 2.1-2).  We introduce the parameters of our specific simulations in \S 2.3.  In \S 2.4 we discuss how we make projections of observables.  We then discuss our results (\S 3), first focusing on how star formation affects the disk and then the stripped tail.  In \S 4 we compare our results to observations and previous simulations.  We discuss the possible effects of our resolution in \S 5.  Finally, we conclude in \S 6 with a summary of our results and predictions for observers.

\section{Methodology}

We use the adaptive mesh refinement (AMR) code {\it Enzo}.   To follow the gas, we employ an adaptive mesh for solving the fluid equations including gravity (Bryan 1999; Norman \& Bryan 1999; O'Shea et al. 2004).  The code begins with a fixed set of static grids and automatically adds refined grids as required in order to resolve important features in the flow.

Our simulated region is 311 kpc on a side with a root grid resolution of $128^3$ cells.   We allow an additional 6 levels of refinement, for a smallest cell size of 38 pc.  We refine the grid based on the local gas mass, such that a cell was flagged for refinement whenever it contained more than about 4900 $\msun$.  We found that these parameters quickly refined most of the galactic disk to 38 pc resolution.  The run also refined much of the wake to a spatial resolution of about 76 pc, and the dense clouds to 38 pc.

The simulation includes radiative cooling using the Sarazin \& White (1987) cooling curve extended to low temperatures as described in Tasker \& Bryan (2006).  To mimic effects that we do not model directly (such as turbulence on scales below the grid scale, UV heating, magnetic field support, or cosmic rays), we cut off the cooling curve at a minimum temperature $T_{\rm min}$ so that the cooling rate is zero below this temperature.  In the simulations described here we use $\tmin = 300$ K, below the threshold for neutral hydrogen formation.  In previous work (Tonnesen \& Bryan 2009; 2010), we have explored the impact of adopting a $\tmin$ value of 8000 K, finding the effect to be relatively small.

\subsection{Star Formation and Feedback Implementation}\label{sec:sf}

Star formation occurred in our finest grid cells (38 pc) when two criteria were met:  1) the gas density in a cell exceeded a critical overdensity (in our runs, this was set to a density of about 3.85 $\times$ 10$^{-25}$ g cm$^{-3}$), and 2) the gas temperature was below 1.1 $\times$ 10$^4$ K.  

The reader may note that this set of criteria is missing two commonly used requirements for star formation (e.g. Cen \& Ostriker 1992):  (1) there is a convergent flow, and (2) the mass in the cell exceeds the Jean's mass.  We chose not to require a convergent flow because we intend to look for star formation in the stripped gas tail and may not be able to resolve the internal structure of the clouds in the tail accurately.  In a previous paper (Tonnesen \& Bryan 2010), we have shown that dense gas can be accelerated to nearly 1000 km s$^{-1}$.   In addition, we have dropped the requirement that the mass in the cell must exceed the Jeans mass because with this condition, our minimum temperature floor could prevent star formation except in the densest cells.  Using our minimum allowed gas density and maximum temperature for star formation, the Jeans mass is 300 times the mass in a cell.  We fulfill the Truelove criterion (1997) using those parameters, but will discuss the limits of our resolution in more detail in Section \ref{sec:starres}.

The implementation of star formation and feedback is explained in detail in Tasker \& Bryan (2006), and our summary here directly reflects that paper (but is included for completeness).  When the gas in a cell meets our requirements for star formation, some of the gas is turned into a star particle.  The mass of the star particle is:

\begin{equation}
m_* = \epsilon\frac{\Delta t}{t_{dyn}}\rho_{gas}\Delta x^3
\end{equation}\label{eq:sfmass}
in which $\epsilon$ is the star formation efficiency, $\Delta t$ is the timestep, $t_{dyn}$ is the local dynamical time, $\rho_{gas}$ is the gas density and $\Delta x$ is the cell size.  The efficiency parameter was chosen to match the Kennicutt-Schmidt relation (as in Tasker \& Bryan 2006).  The star formation and feedback parameters we use are given in Table \ref{tableSF}.  

If the above requirements are met and the resulting star particle will have a mass above a minimum mass, m$_{*min}$, it is formed.  This mass is chosen so that a large number of small star particles will not slow down the simulation.  However, if the star particle would have a smaller mass, the probability that it will form is equal to the ratio of the mass of the projected star particle to m$_{*min}$.  If the star particle is then formed, its mass is the minimum of m$_{*min}$ and 80\% of the mass in the gas cell.  Thus, the probability of forming stars in any individual cell is low, but this algorithm still produces stars at the specified rate.  This is done by keeping track of the amount of mass in cells that fulfilled all of the star formation criteria except the minimum mass requirement.  When the mass of the unformed stars reaches the minimum mass, star particles will form even if they are below the specified minimum mass.  In order to make sure that we were not missing any stars formed in the larger volume of the stripped tail, we allowed for a smaller minimum mass in our simulation with ram pressure stripping.  As we will show, this lower m$_{*min}$ does not change the star formation in the disk, but we did find that the lower minimum was necessary to form the correct amount of stars in the tail.

\begin{table}
\begin{center}
\begin{tabular}{c | c}
\hline
Variable & Value\\
\hline
$\rho_{\rm min}$ & 3.85 $\times$ $10^{-25}$ g cm$^{-3}$ \\
T$_{\rm max}$ & 1.1 $\times$ 10$^4$ K \\
$\epsilon$ & 0.5\% \\
m$_{* \rm min}$ & 10$^4$ M$_\odot$ (SFNW) 10$^2$ M$_\odot$ (SFW)\\
$\epsilon_{\rm SN}$ & 10$^{-5}$\\
\hline
\end{tabular}
\end{center}
\caption{Star Formation and Feedback Parameters\label{tableSF}}
\end{table}

We also include stellar feedback from Type II supernova explosions.  Not only may this be important for regulating star formation in the galactic disk (e.g. Tasker \& Bryan 2006; Robertson et al. 2004), but feedback from star formation in a stripped gas tail could enhance the rate at which stripped gas mixes with the ICM.  Star formation in a molecular cloud is likely to be spread out over a dynamical time, and so in order to calculate the timeline of feedback, stars in a particle are considered to form according to the relation:
\begin{equation}
m_{star}(t) = m_* \int^t_{t_{\rm SF}} \frac{(t-t_{\rm SF})}{\tau^2} \exp{ \frac{-(t-t_{\rm SF})}{\tau}} dt,
\end{equation}
where $t_{\rm SF}$ is the time at which the star particle was formed and $\tau$ = max($t_{\rm dyn}$, 10 Myr).  As in Tasker \& Bryan (2006), over a time period of a few $\tau$, 10$^{-5}$ of the rest-mass energy of the stars is added to the thermal energy of the gas in the cell in which the star particle has been created.  This corresponds to approximately 56 solar masses of stars formed for each $10^{51}$ erg SN.

\subsection{The Galaxy}

Our galaxy is placed at a position corresponding to (155.5,155.5,68.42) kpc from the corner of our cubical 311 kpc computational volume, so that we can follow the stripped gas for more than 200 kpc.  The galaxy remains stationary throughout the runs.  The ICM wind flows along the z-axis in the positive direction, with the lower z boundary set for inflow and upper z boundary set as outflow.    The x and y boundaries are set to outflow in all three cases.

We chose to model a massive spiral galaxy with a flat rotation curve of 200 km s$^{-1}$.  It consists of a gas disk that is followed using the adaptive mesh refinement algorithm (including self-gravity of the gas and any newly formed stars), as well as the static potentials of the (pre-existing) stellar disk, stellar bulge, and dark matter halo.  We directly follow Roediger \& Br\"uggen (2006) in our modeling of the stellar and dark matter potential and gas disk.  In particular, we model the stellar disk using a Plummer-Kuzmin disk (see Miyamoto \& Nagai 1975), the stellar bulge using a spherical Hernquist profile (Hernquist 1993), and the dark matter halo using the spherical model of Burkert (1995).  This dark matter halo model is compatible with observed rotation curves (Burkert 1995; Trachternach et al. 2008).  The equation for the analytic potential is in Mori \& Burkert (2000).   We describe our disk in detail in Tonnesen \& Bryan (2009, 2010).   Briefly, our stellar disk has a radial scale length of 4 kpc, a vertical scale length of 0.25 kpc and a total mass of 10$^{11}$ M$_{\odot}$; the stellar bulge has a scale length of 0.4 kpc and a total mass of 10$^{10}$ M$_{\odot}$; and the dark matter halo has a scale radius of 23 kpc and a central density of $3.8 \times 10^{-25}$ g cm$^{-3}$.  The gas disk has about 10\% of the mass in the stellar disk, and radial and vertical scales of 7 kpc and 0.4 kpc, respectively.

To identify gas that has been stripped from the galaxy we also follow a passive tracer that is initially set to 1.0 inside the galaxy and $10^{-10}$ outside.  In the following analysis, we will use a minimum tracer fraction of 25\% to find gas stripped from the galaxy (our conclusions do not change if we use 10\% instead).

\subsection{The Simulations}\label{sec:sims}

All three of the galaxies we discuss in this paper initially evolve in a static, high-pressure medium with $\rho=$ 9.152 $\times$ 10$^{-29}$ g cm$^{-3}$ and $T = $ 4.15 $\times$ 10$^6$ K, to allow cool, dense gas to form in the galaxy.  This naturally generates a multiphase ISM (see Tasker \& Bryan (2006) and Tonnesen \& Bryan (2009) for more discussion of the ISM properties).  

In our simulation without star formation, after 155 Myrs we reset the boundary conditions to generate a constant ICM inflow along the inner z-axis, which is always face-on to the galaxy.   We chose this time so that the galaxy would have formed high density gas clouds ($\rho > 10^{-23}$ g cm$^{-3}$) by the time the wind hits the galaxy (190 Myr after the start of the simulation).  In our comparison case with star formation, we delay the onset of the wind by about 18 Myr in order to allow the galaxy to evolve for more than 200 Myr before the wind hits the disk.  We do this for two reasons:  first, because Tasker \& Bryan (2006) found that the star formation rate in the disk settles to a relatively constant value after about 200 Myr of star formation, and second, because at that point in our simulations the azimuthally-averaged relation between gas surface density and SFR surface density agrees with that found in Kennicutt (1989, 1998).  While we could have chosen to wait longer, this change would have no qualitative effect on our conclusions.  

In this paper we discuss three simulations.  All of these runs have the same initial conditions (same galaxy density profiles evolving in a static ICM with $\rho=$ 9.152 $\times$ 10$^{-29}$ g cm$^{-3}$ and $T = $ 4.15 $\times$ 10$^6$ K and a cooling curve following Sarazin \& White (1987) extended to $\tmin = 300$ K).  Two of these simulations include star formation, SFNW and SFW.  SFNW evolves in a static ICM, while SFW initially evolves in a static ICM and is later stripped by a higher-density ICM wind.  The final simulation we discuss in this paper is NSFW, which is the same simulation as SFW without star formation.  NSFW is the $\tmin = 300$ K case discussed in our earlier paper, Tonnesen \& Bryan (2010).  In both wind cases,  $P_{\rm ram} = \rho v^2_{\rm ICM} = 6.4 \times 10^{-12}$ dynes cm$^{-2}$, and $v_{ICM} = 1413$ km s$^{-1}$.  The ICM wind has a T = 4 $\times$ 10$^7$ K and $\rho$ = 3.2 $\times$ 10$^{-28}$ g cm$^{-3}$.

\subsection{Projections}\label{sec:projection}

{\it Enzo} outputs the density and temperature of the gas in each cell.  To transform these values into \ion{H}{I} column density and H$\alpha$ intensity, we used Cloudy, version 08.00 of the code, last described by Ferland et al. (1998).  Using a grid of temperatures and densities, we calculated the hydrogen neutral fraction and H$\alpha$ emissivity.   In the Cloudy calculation, we included cosmic microwave background radiation, the cosmic ray background, bremsstrahlung radiation from the ICM and the 2005 version of the Haardt \& Madau (2001) $z=0$ metagalactic continuum, as implemented by Cloudy.  We found that including a local interstellar radiation field emission resulted in somewhat lower amounts of neutral gas.  Since much of our gas is very distant from the galaxy, we decided not to include this radiation.  We also found that removing bremsstrahlung radiation did not significantly change any of the values we considered. 

We chose to calculate the neutral fraction and H$\alpha$ emissivity (from collisionally heated diffuse gas; emission from HII regions will be dealt with separately) for a thin plane-parallel gas cloud of width 100 pc.  We selected this width because it corresponds roughly to the cell size of most of the gas in the highly resolved tails, and accounts approximately for radiative transfer effects.  If we assumed the radiative thin limit, it would slightly decrease the amount of \ion{H}{I} we predict, and slightly increase the H$\alpha$ emission for dense, low-temperature gas.  Ideally, we would include the radiation field with radiative transfer directly in the simulation, but this is not yet feasible (and only has a slight impact on the dynamics); instead we post-process these results to get reasonable predictions for the ionization fraction and H$\alpha$ emissivity (see Furlanetto et al. 2005 for a discussion of various approaches in the context of Ly$\alpha$ emission).  For a more detailed discussion, we refer the reader to Tonnesen \& Bryan (2010). 

\section{Results}

\subsection{Star Formation in the Galactic Disk}

\begin{figure}
\includegraphics{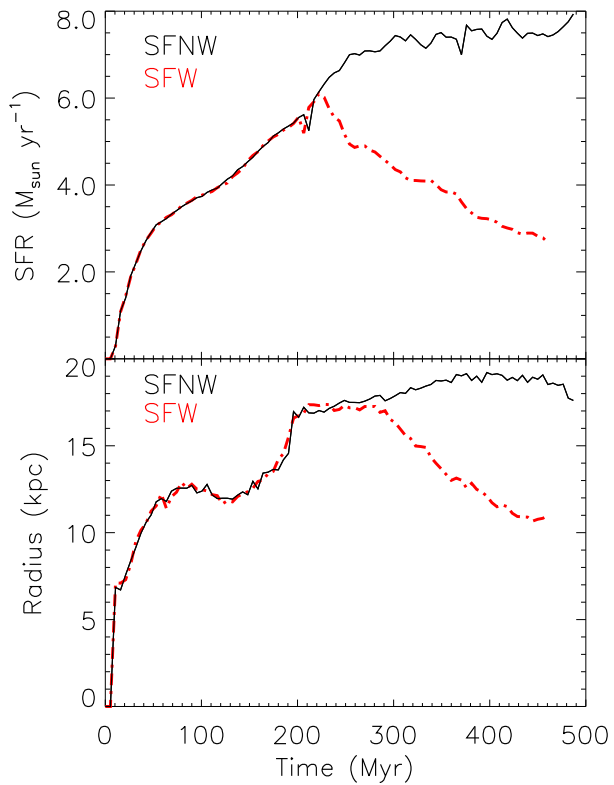}
\caption{The top panel plots star formation rate (SFR) as a function of time in the simulated galaxy with no wind (SFNW, black line) and in the galaxy that is hit by an ICM wind after about 210 Myr (SFW, red line).  About 10 Myr after the wind hits the disk, the SFR of the galaxy begins quickly decreasing, and continues to decrease throughout our run, as the disk gas is stripped.  The bottom panel plots the disk radius (in kpc) that includes 95\% of the new stars formed as a function of time for the same runs as in Figure~\ref{fig-SFR}.  Once the wind hits the SFW galaxy (red), this outer star forming radius decreases, but not as quickly as the SFR (Figure \ref{fig-SFR}), probably because some dense clouds in the disk cannot be stripped by the wind, so instead form stars.}\label{fig-SFR}
\end{figure}

We will first compare the star formation in the galactic disks of the SFNW and SFW cases to determine if and how ram pressure stripping affects disk star formation.  We define the disk to extend 2 kpc from the the central disk plane, which includes all of the star formation in the SFNW run.  First, in the top panel of Figure \ref{fig-SFR} we plot the star formation rate (SFR) as a function of time for the SFNW and SFW runs.  For the first 220 Myr the SFRs are nearly identical.  The higher m$_{* \rm min}$ in SFNW only makes the SFR slightly less smooth over time, and does not affect the agreement of SFNW and SFW.  Shortly after the ICM wind hits the SFW galaxy disk, its SFR drops precipitously.  This figure clearly shows that ram pressure stripping quickly lowers the SFR and does not induce even a short-lived burst of star formation (at least at the level of ram pressure modeled here).  While our exact star formation recipe results in a SFR in our models that is high for an isolated Milky-Way sized spiral galaxy, we emphasize that it is the comparison between SFNW and SFW that is important in this work.

In the bottom panel of Figure \ref{fig-SFR} we plot the radius including 95\% of the new stars formed in the disk against time.   As in the panel above, we find that the two cases are nearly identical for the first 200 Myr.  However, it takes $\sim$ 70 Myr longer than for the SFR (until $\sim$290 Myr) for the star-formation radii to diverge.  This is likely because dense clouds have formed up to 17 kpc from the disk centre that cannot be instantly stripped by the ICM wind.  It is only after these clouds have formed stars or been ablated --- had their own edges stripped by the wind until they are destroyed or are of low enough density to be removed from the disk by the ICM wind --- that the star-formation radius drastically drops.   

\begin{figure}
\begin{center}
\includegraphics{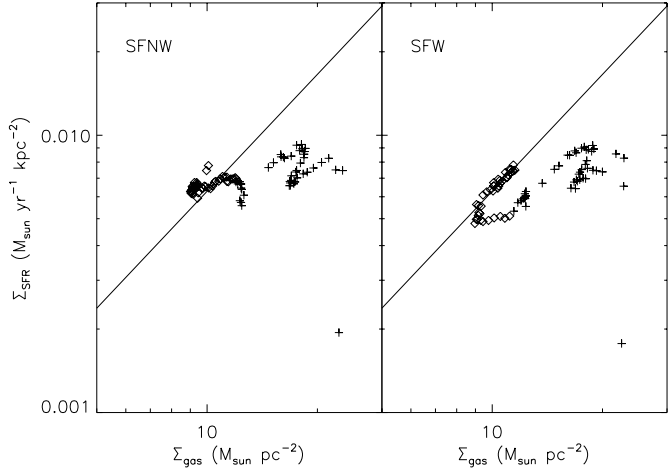}
\caption{The Kennicutt-Schmidt relation of the two simulated galaxies.  For each timestep we calculate the gas and SFR surface density within the radius containing 95\% of the new star formation for that timestep.  Plus symbols are used for the first 250 Myr, before the disk has completely settled, and then diamonds are used thereafter.  At early times, the star formation relation lies below the line denoting the observed Kennicutt-Schmidt relation (Kennicutt 1989), while at late times the galaxies lie very close to the Kennicutt-Schmidt Law. }\label{fig-schmidtlaw}
\end{center}
\end{figure}

In the two previous figures we focused on the total star formation rate.  Now we will look at how star formation rate relates to the gas surface density.  Even though the star formation rate does not spike when the wind hits the galaxy, it could be high relative to the surface density of the gas remaining in the disk.  In Figure \ref{fig-schmidtlaw} we plot the Schmidt-Kennicutt relationship of each galaxy for each timestep (Schmidt 1959; Kennicutt 1989).  We focus on the outputs at times later than 250 Myr (shown as diamonds in this figure), as this is both when the SFR becomes constant in the SFNW case (see Fig. ~\ref{fig-SFR}), and also after the SFW galaxy begins being ram pressure stripped.  Limiting ourselves just to those points after 250 Myr, we can see that the Schmidt-Kennicutt relationship closely follows the observed relation in both cases, plotted as the solid line (2.5 $\times$ 10$^{-4}$ ($\Sigma_{gas}$/ 1 M$_\odot$ pc$^{-2}$)$^{1.4}$).  At the latest times in the SFNW galaxy, the galaxy evolves only very slowly in gas surface density or SFR surface density.  The SFW galaxy has a slightly increasing gas surface density with time because the outer, lower-density regions are being stripped.

\begin{figure}
\includegraphics{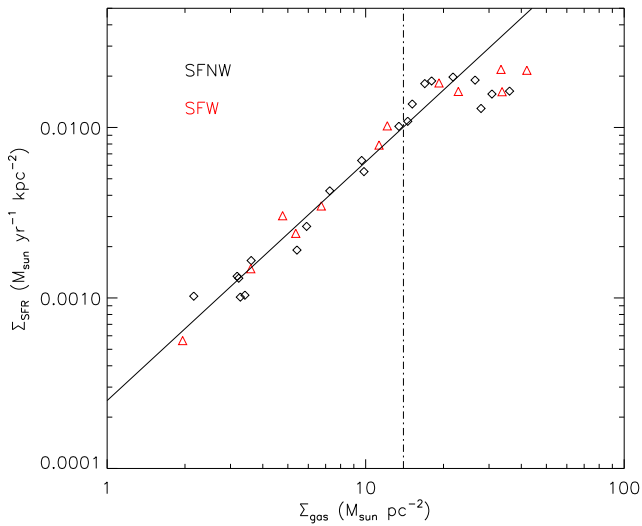}
\caption{The star formation-surface density relation computed in rings with widths of 1 kpc.  The SFNW run is shown with black diamonds and SFW is the red triangles, both shown at a time 250 Myr after wind hits disk (460 Myr into the simulation).  The solid line is the empirical Kennicutt relation (2.5 $\times$ 10$^{-4}$ ($\Sigma_{gas}$/ 1 M$_\odot$ pc$^{-2}$)$^{1.4}$) (Kennicutt 1989), and the dash-dotted line denotes the gas surface density at which Leroy et al. (2008) found the local star formation efficiency flattens.}\label{fig-localschmidt}
\end{figure}

In Figure \ref{fig-localschmidt} we plot a ``local" Schmidt Law--the SFR surface density against the gas surface density in rings with 1 kpc width.  This relationship is plotted for a single time snapshot--460 Myr into the simulation, or 250 Myr after the wind hits the SFW galaxy.  We choose this late time because it maximizes the difference between the SFR and the star-formation radius of the SFW and SFNW galaxies (as seen in Figure \ref{fig-SFR}).  However, clearly the local relationships between SFR surface density and gas surface density are very similar in both runs.  They even seem to flatten at about the same gas surface density, $\sim$20 M$_\odot$ pc$^{-2}$, which is in remarkably good agreement with the gas surface density at which the local SF efficiency flattens as observationally found by Leroy et al. (2008) (14 $\pm$ 6 M$_\odot$ pc$^{-2}$; see their Figure 5).  The most notable difference between our two runs is that the SFW galaxy has fewer points than the SFNW galaxy.  This could be because either the gas density is zero in a 1 kpc ring in the galaxy, or the SFR surface density is zero.  In fact, both the gas density and SFR surface density are very low outside of a radius of about 12 kpc, continuing to hold to a correlation between gas and SFR surface density.

\begin{figure}
\includegraphics{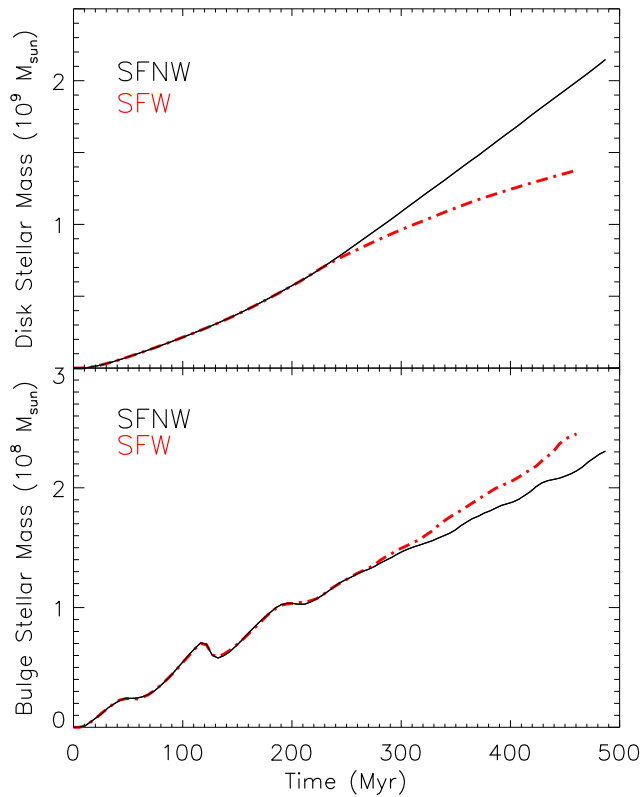}
\caption{The top panel shows the total amount of newly formed stellar mass in the disk (i.e. within 2 kpc of the disk plane), as a function of time.  Shortly after the wind hits the galaxy on the right, the star formation rate decreases, and by the end of the SFW run, it has about 7 $\times$ 10$^8$ M$_\odot$ less stellar mass in the disk than the SFNW run.  The bottom panel plots mass of newly formed bulge stars, as a function of time.  If we focus only on the bulge -- all stars within 3.4 kpc of the galaxy centre -- we see that ram pressure does lead to more stars in the bulge.  However, the difference does not significantly change the B/T ratio, which begins at 0.1 with M$_{Bulge}$ = 10$^{10}$ M$_\odot$ and M$_{Disk}$ = 10$^{11}$ M$_\odot$.}\label{fig-smassdisk}
\end{figure}

Thus far we have found that including a ram pressure stripping wind decreases the total SFR and focuses the SF towards the centre of the galaxy (Figure \ref{fig-SFR}), but only slightly increases the gas surface density in the disk and does not change the relationship between gas surface density and SFR surface density (Figures \ref{fig-schmidtlaw} and \ref{fig-localschmidt}).  We finally consider the total amount of newly formed stellar mass in the disk and in the bulge of the galaxy.  As shown in the top panel of Figure~\ref{fig-smassdisk}, the total stellar mass formed in the disk, since the beginning of the simulation, is nearly identical in SFNW and SFW for the first 200 Myr, but once the wind hits the disk, the two lines begin to diverge.  By the end of the SFW simulation, the SFW galaxy has about 7 $\times$ 10$^8$ M$_\odot$ less stellar mass in its disk.  

If we focus only on the (newly formed) bulge stars, as defined by all of the stars formed since the simulation began in a sphere with a 3.4 kpc radius from the galaxy centre (this includes 80\% of the mass in the spherical Hernquist bulge we initially used to determine our galaxy potential), we find that including the wind leads to more stars in the galactic bulge, as shown in the bottom panel of Figure~\ref{fig-smassdisk}.  As we have discussed in Tonnesen \& Bryan (2009), we find that gas clouds that are not stripped are able to spiral towards the centre of the disk (initially seen by Schulz \& Struck 2001).  It is this inflow of gas within the disk that adds most of the stars to the bulge (rather than stellar fallback).  

Our bulge-to-total ratio of new stars is 0.1 in the SFNW galaxy and 0.2 in the SFW galaxy.  However, this galaxy initially had a stellar disk mass of 10$^{11}$ M$_\odot$ and a stellar bulge mass of 10$^{10}$ M$_\odot$, so the the new stars change the bulge-to-total ratio of stellar mass by at most a few percent.    In summary, we find that ram pressure stripping lowers the SFR of a galaxy without an initial burst, despite the fact that our stripping ICM has a higher pressure than the static ICM.  The relationship between gas density and SFR is similar in both simulations, with a slight increase in the total gas surface density for the SFW run, as would be expected from both the ram pressure and the higher-pressure surrounding ICM.  Finally, although ram pressure does cause SFW to form more stars in the bulge and less stars in the disk than SFNW, it is not enough to overcome the initial mass profile of the galaxy.  The galaxy would have to have about 2 orders of magnitude less stellar mass for the difference in star formation to have a significant effect on the bulge-to-total mass ratio.

\subsection{Gas in the Galactic Disk}

\begin{figure}
\includegraphics{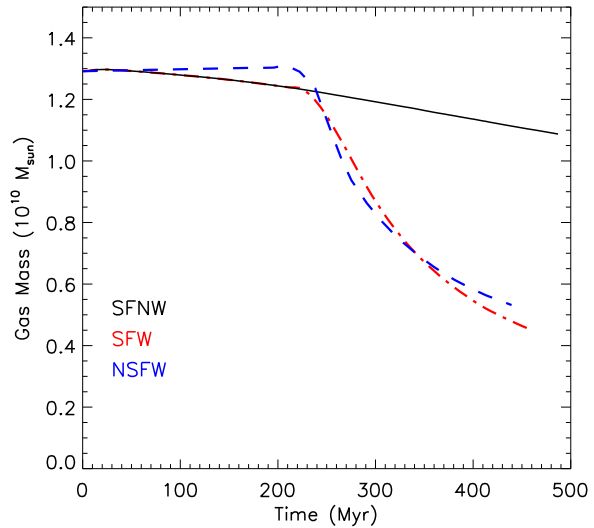}
\caption{The amount of gas in the disk in all three simulated galaxies.  The SFW galaxy has less gas than the NSFW galaxy after being stripped for 250 Myr, but only by 10$^{9}$ M$_\odot$.  This is less than the 1.4 $\times$ 10$^9$ M$_\odot$ of stars that form throughout the SFW simulation.}\label{fig-gasmass}
\end{figure}

We will now examine if including star formation and feedback affects the remaining gas disk of a ram pressure-stripped galaxy.  First we plot the amount of gas in the disk as a function of time for all three simulations in Figure \ref{fig-gasmass}.  Gas is counted as being in the disk if it has a tracer fraction of more than 0.6 and is within 2 kpc of the disk central plane.  Focusing on SFNW (the black solid line), we see that including star formation results in disk gas being used to form new stars throughout the simulation.  Comparing the two ram pressure stripped galaxies (SFW: red dash-dotted line and NSFW: blue dashed line), we see that after 250 Myr of stripping (the end of each line), the galaxies lose very similar amounts of gas.  The SFW run has about 10$^9$ M$_\odot$ less gas left in the disk than the NSFW galaxy, which is less than the amount of gas that formed stars (1.4 $\times$ 10$^9$ M$_\odot$), so 4 $\times$ 10$^8$ M$_\odot$ less gas was actually removed from SFW than from NSFW by the ICM wind.  We conjecture that this is because dense gas clouds near the outer edges of the disk formed stars in SFW rather than being ablated and eventually stripped by the ICM wind.

\begin{figure*}
\includegraphics{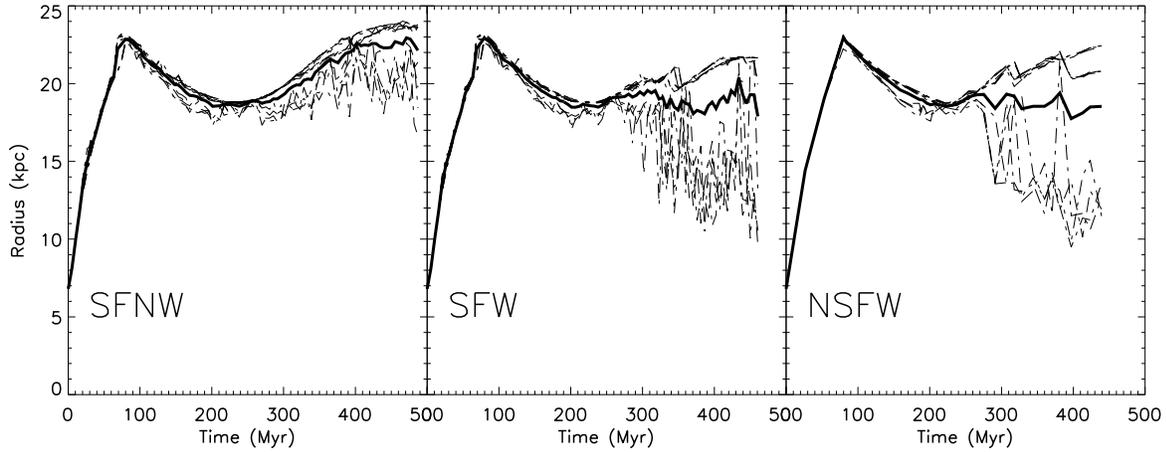}
\caption{The maximum radius of gas with a density above 10$^{-24}$ g cm$^{-3}$ as a function of time.  Dash-dotted lines show this radius for each of twelve wedges of the disk, while the solid line is the mean of the wedges.  From left to right, the panels show SFNW, SFW, and NSFW.  Being stripped by the ICM wind lowers the galaxy radius, but including star formation does not have much affect on the radius of the dense gas that remains in the disk (at least for 250 Myr of stripping).}\label{fig-gasrad}
\end{figure*}

This picture of outer gas clouds being either stripped (NSFW) or forming stars (SFW) also agrees with Figure \ref{fig-gasrad}, where we plot the radius of dense gas in the disk in all three simulations (see also Tonnesen \& Bryan 2008).  This is the radius of gas with a density above 10$^{-24}$ g cm$^{-3}$.  For each timestep we consider twelve wedge-shaped sections of the disk and measure the largest radius at which there is gas with a density above 10$^{-24}$ g cm$^{-3}$.  Each of these individual measurements are shown as dash-dotted lines.  We also plot the mean radius of all wedges against time as the thick solid line.  In all three cases, we see the collapse of the disk gas into dense clouds as the early increase in the radius of this dense gas.  The ram pressure stripping wind affects both the disk with and the disk without  star formation in a similar fashion: the gas disks have similar ranges of measured radii and a similar mean radius of about 18-19 kpc.  

\begin{figure*}
\includegraphics{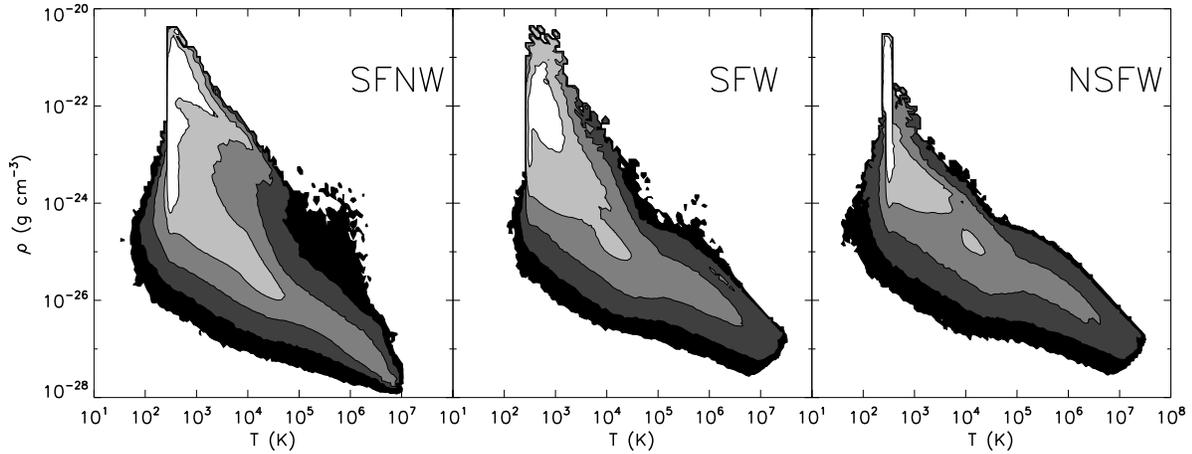}
\caption{Contours of gas mass in the disk (defined as gas with a tracer fraction of more than 0.6 and a height above the disk of less than 2 kpc), as a function of gas density and temperature.  These are all shown 250 Myr after the wind has hit the galaxy (or simply 460 Myr into the simulation with no ICM wind).  Including star formation and feedback allows gas in the disk to have lower densities and higher temperatures which may make gas in the disk with star formation easier to strip.
}\label{fig-rhotdisk}
\end{figure*}

We next consider how star formation affects the density and temperature distribution of gas in the galactic disk.  In Figure \ref{fig-rhotdisk}, we show contours of gas mass in the disk, as a function of gas density and temperature.  These are all 250 Myr after the wind has hit the galaxy (or simply 460 Myr into the SFNW simulation).  Including star formation and feedback spreads the distribution of high density ($\rho >$ 10$^{-23}$ g cm$^{-3}$) gas in the disk to include lower densities and higher temperatures.  This may lower the surface density of gas in the galaxies with star formation, making it easier to strip.  

\begin{figure*}
\includegraphics{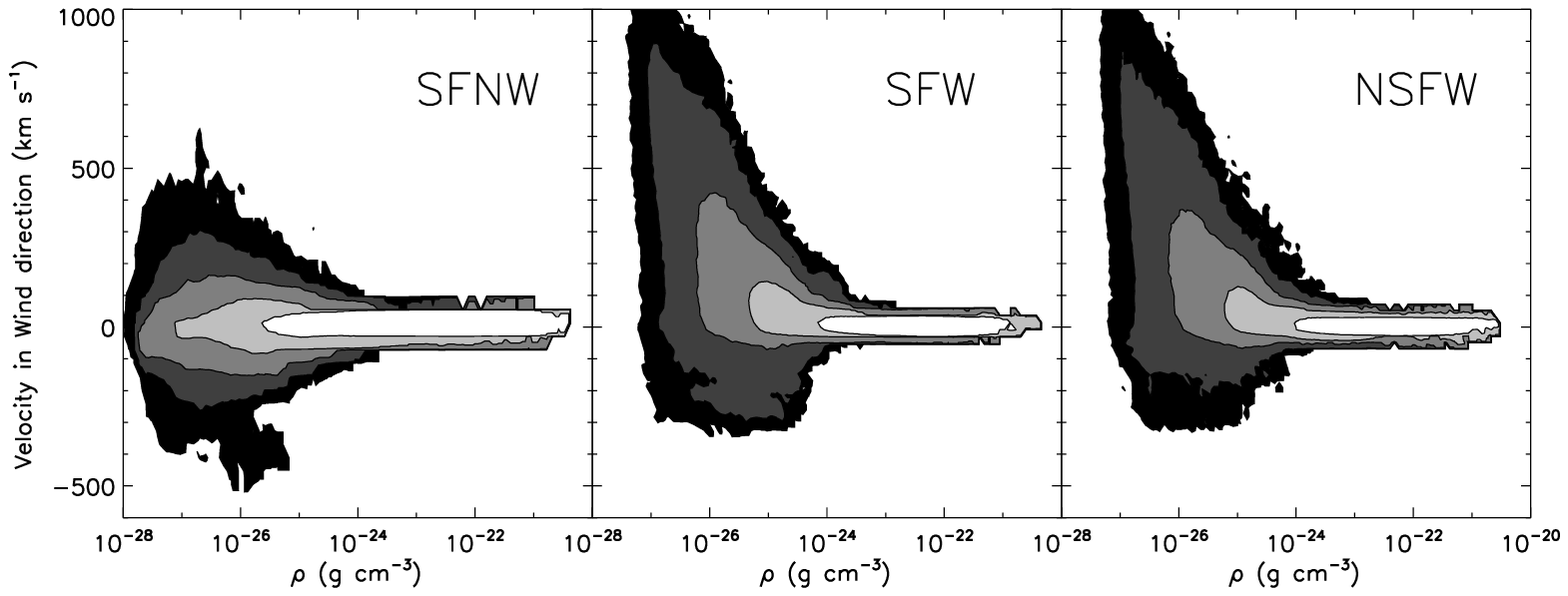}
\caption{Contours of gas mass in the disk as a function of velocity in the wind direction and gas density, 250 Myr after the wind has hit the galaxy.  Including star formation and feedback causes gas in the disk to have negative and positive velocities.  In addition, at any density, the run that includes star formation has slightly more gas at higher velocities. }\label{fig-rhovzdisk}
\end{figure*}

Finally we consider the velocity structure of the gas in the disk.  In Figure \ref{fig-rhovzdisk}, we plot contours of gas mass in the disk (defined as before), as a function of density and gas velocity in the wind direction, for the same time as in Figure~\ref{fig-rhotdisk}.  First, we note that the velocity spread in the SFNW run indicates the density and velocity of the disk gas that is accelerated due to the inclusion of thermal feedback from supernovae (i.e. only gas with $\rho \leq 10^{-24}$ g cm$^{-3}$ is affected).  

Gas accelerated by this process likely explains most of the differences between the SFW and NSFW cases.  The most obvious difference is in the gas with negative velocities -- while in the NSFW runs, most of the gas with negative velocities is likely the fallback of stripped gas into the central region of the disk, in the SFW simulation a portion of the gas is also accelerated due to feedback.  A smaller effect is seen in the positive velocity gas -- the same density gas can have a higher positive velocity in the SFW case than in the NSFW case.  

However, this figure tells us that the gas being stripped (with large positive velocities) has densities less than 10$^{-23}$ g cm$^{-3}$, so the differences that star formation and feedback cause in the density and temperature distribution of disk gas (Figure \ref{fig-rhotdisk}) do not strongly affect gas that will be stripped.  Although we do not carry out a series of runs with higher ram pressures that also include star formation (due to prohibitively long run times), we have analyzed simulations with higher ram pressures (but without star formation) in Tonnesen et al. (2011).  In that case, with a ram pressure of 4 $\times$ 10$^{11}$ dynes cm$^{-2}$, gas with densities as high as 6 $\times$ 10$^{-23}$ g cm$^{-3}$ can be stripped.  We predict that star formation would result in that galaxy being stripped more quickly than in our simulations with no star formation.

Including star formation only has a slight effect ($\sim$10\%) on the amount of gas left in the disk (Figure \ref{fig-gasmass}) and has very little effect on the residual size of the unstripped gas disk (Figure \ref{fig-gasrad}).  We do find noticeable differences in the density-temperature and z-velocity distributions of the gas in the disk, likely due to feedback from star formation.

\subsection{Gas in the Stripped Tail}

\begin{figure*}
\includegraphics{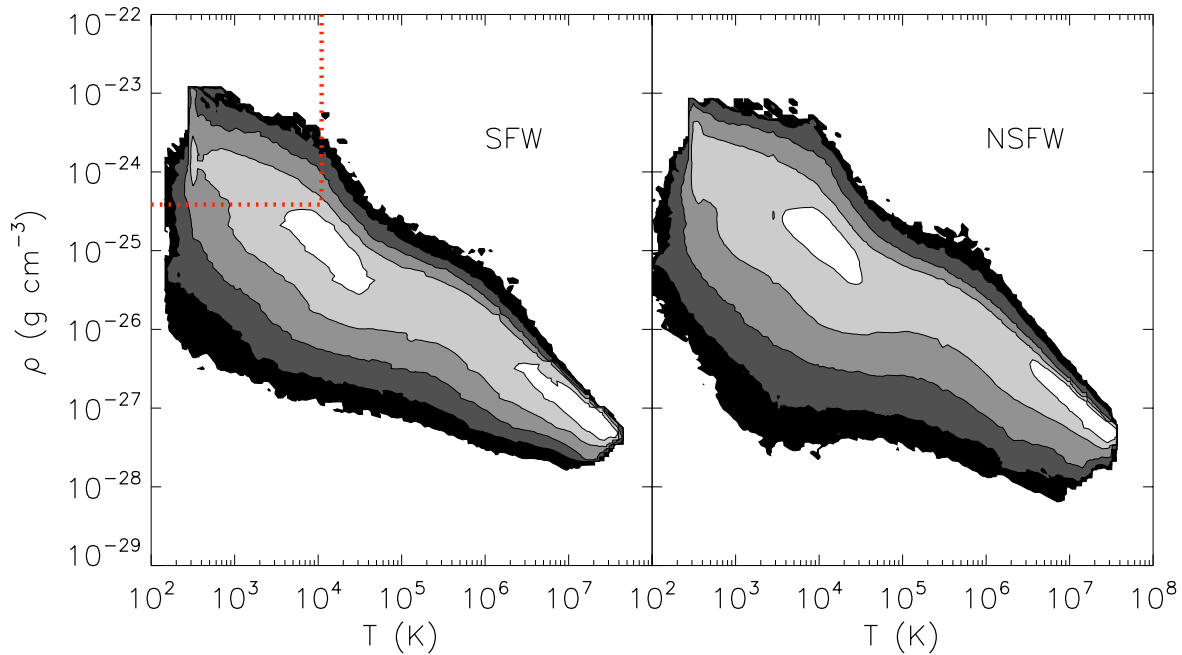}
\caption{Contours of gas mass in the tail (tracer fraction of more than 0.25 and height greater than 10 kpc), as a function of gas density and temperature, both at a time 250 Myr after the wind has hit the galaxy.  Including SF and feedback has a minimal effect on the $\rho$-T distribution of the tail.  The dotted red lines denote the minimum mass and maximum temperature necessary for stars to form from a gas cell.}\label{fig-rhot}
\end{figure*}

\begin{figure*}
\includegraphics{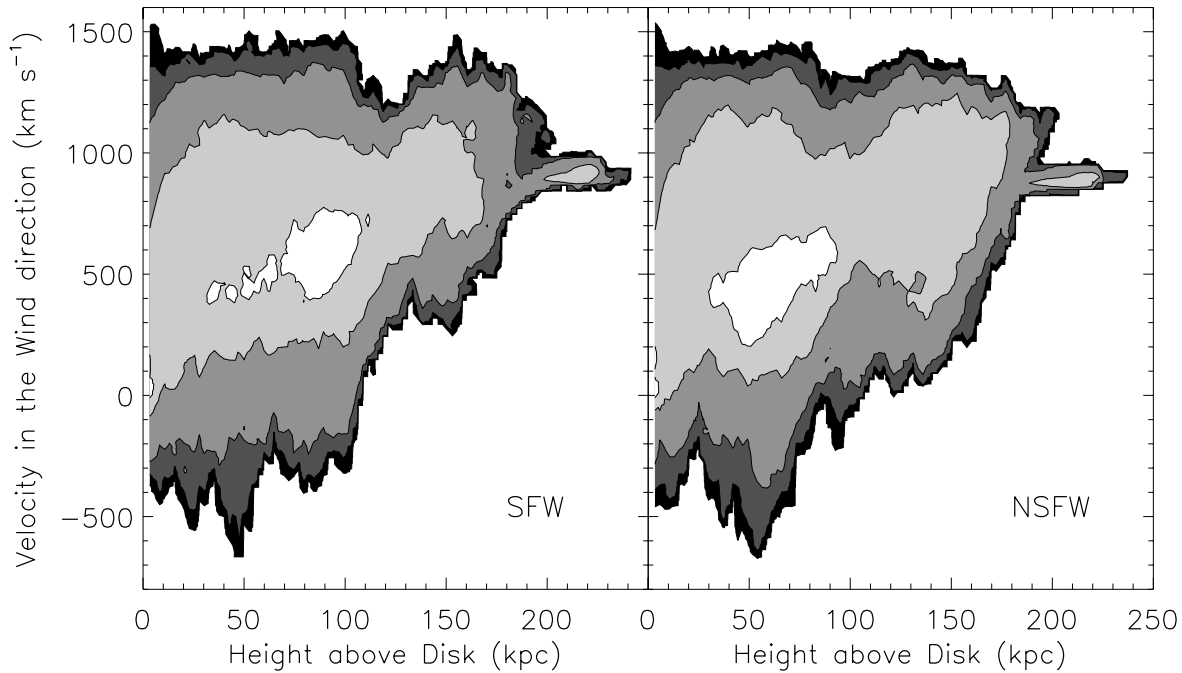}
\caption{Contours of gas mass in the tail as a function of gas velocity in the wind direction and height above the disk.  These are both 250 Myr after the wind has hit the galaxy.  Including star formation results in the bulk of the stripped gas being farther above the disk, although the range of z-velocities is very similar both cases.}\label{fig-rhovz}
\end{figure*}

\begin{figure*}
\includegraphics{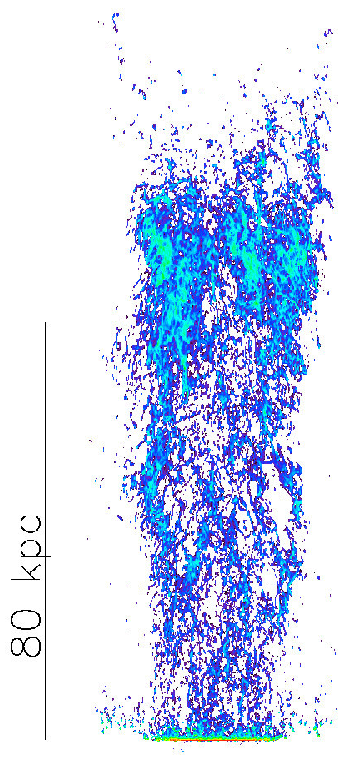}
\includegraphics{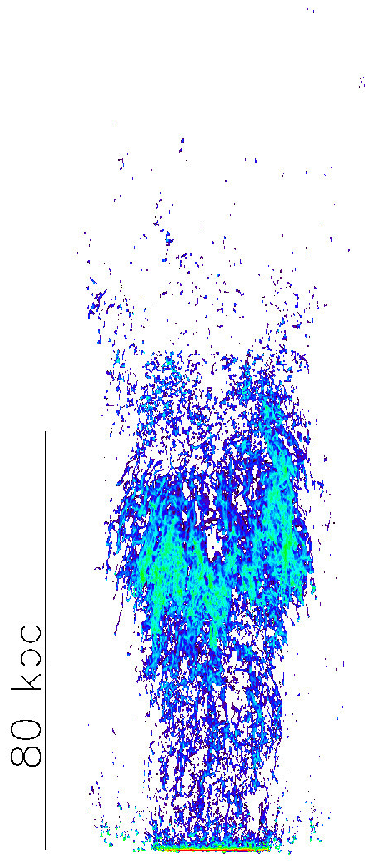}
\includegraphics{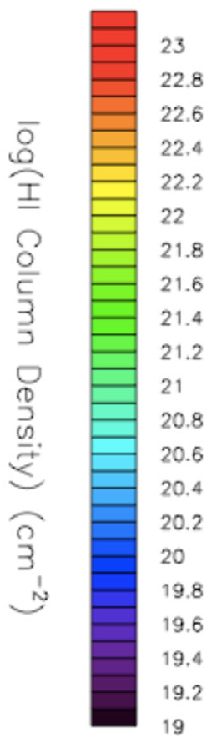}\\
\includegraphics{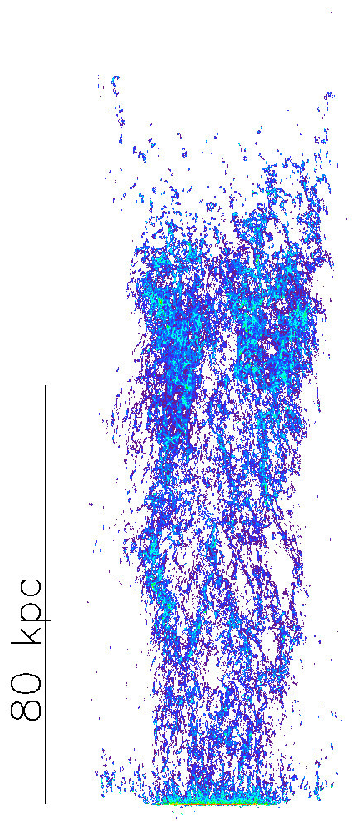}
\includegraphics{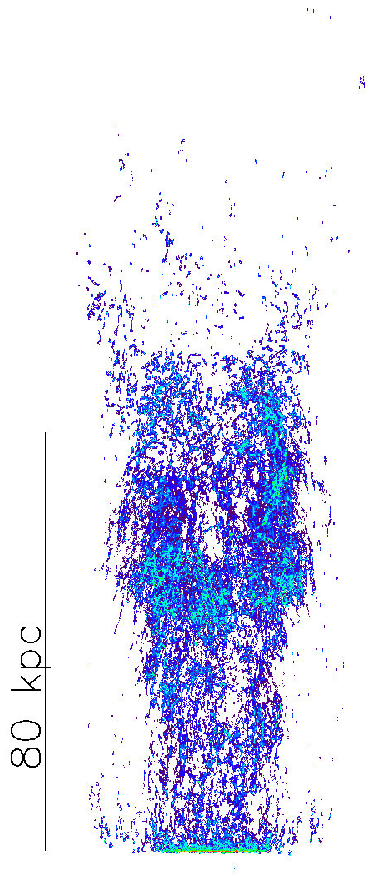}
\includegraphics{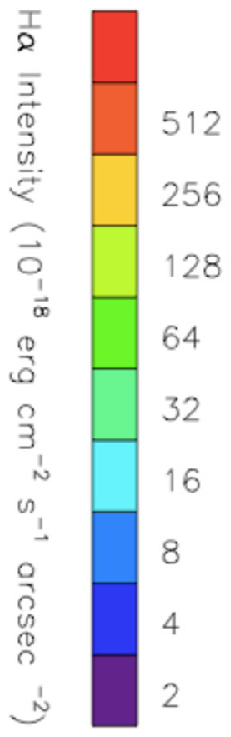}
\caption{Projections of HI column density (top row) and H$\alpha$ intensity (bottom row).  The galaxy with star formation and feedback (SFW) is on the left, and without (NSFW) is on the right.  Including star formation results in slightly longer tails.}
\label{fig-projs}
\end{figure*}

We turn now to the gas in the stripped tail.  First, as with the disk gas, we plot contours of gas mass as a function of density and temperature in Figure \ref{fig-rhot}.  This figure is a snapshot 250 Myr after the wind has hit the disk, including all of the gas more than 10 kpc above the disk with a tracer fraction above 0.25 (i.e only gas that originated in the disk).  It is immediately clear that the gas density and temperature distributions in the tail are very similar whether or not the simulation includes star formation.  In the SFW panel we have denoted the range of temperatures and densities at which gas may form stars.  The NSFW contours reach somewhat lower densities at low temperatures.  In Figure \ref{fig-rhovz} we plot contours of gas mass as a function of velocity in the wind direction and height above the disk.  Once again, the distributions are very similar, although on closer inspection there is a slight difference in the placement of the highest contour (see also Figure~\ref{fig-projs}), and the SFW tail has a narrower velocity distribution at large distances from the disk.  

We can see how the similarity in density, temperature, and velocity plays out in observables, specifically \ion{H}{I} column density and H$\alpha$ emission.  In Figure \ref{fig-projs} we display the projections of the SFW run on the left and the NSFW run on the right.  For these plots, we restrict ourselves to diffuse H$\alpha$ emission, neglecting the contribution from HII regions (see the next section for the stellar contribution).  Unlike the previous figures, here the tails look somewhat different, with the SFW tail having much of its dense gas farther from the disk than the NSFW tail.  The small difference in the highest contour of Figure \ref{fig-rhovz} has resulted in a significantly different distribution of bright \ion{H}{I} and H$\alpha$ emission.  

These differences in the distribution of the stripped gas are due to feedback near the disk.  Thermal feedback results in gas outflows above and below the disk.  This means that some gas that will be removed from the galaxy is already moving away from the galaxy in the direction of the ICM wind.  This is why dense gas is farther from the disk in the SFW simulation than in the NSFW simulation.  There is not much star formation in the stripped tail, as we will show below, so it does not have such a large effect on the morphology of the tail.  

As we would expect from the similarity between SFW and NSFW in Figure \ref{fig-rhot}, the range of \ion{H}{I} column densities and H$\alpha$ intensities are the same in the two tails.  The total emission from the tails is also very similar.  Including star formation results in slightly less \ion{H}{I} gas (20\% less), possibly because some of the most dense gas turns into stars.  The (diffuse) H$\alpha$ emission is only 2\% higher when including star formation and thermal feedback.  Although we do not show a projection here because the X-ray brightness is too low to be observed, including star formation reduces the X-ray luminosity by only a small amount, 13\%.

\subsection{Star Formation in the Stripped Tail}

\begin{figure}
\includegraphics{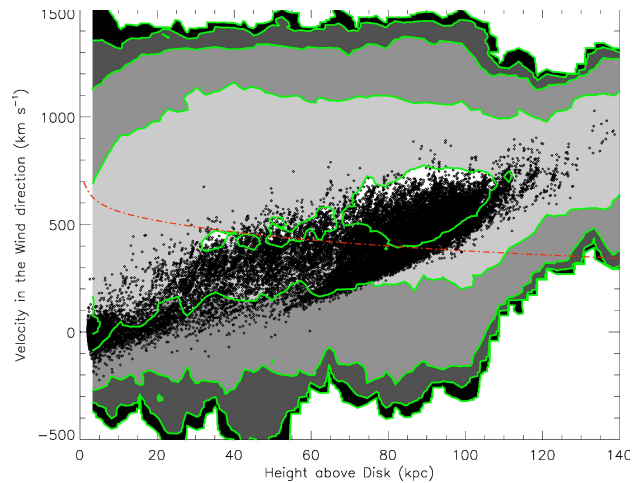}
\caption{The z-velocity and height above the plane of the star particles are shown as diamonds (showing only those with heights above 1 kpc) are overplotted on the SFW gas contours from Figure \ref{fig-rhovz}.  The edges of the gas contours are overplotted in green for clarity.  The stars begin with the velocity of the gas in the tail from which they are formed, and then slow down as they are no longer accelerated by the wind.}\label{fig-starpos}
\end{figure}

Finally, we consider star formation in the tail.  In Figure \ref{fig-starpos} we plot the z-velocity of the tail stars against the height above the disk for a single output 250 Myr after the wind has hit the disk.  We plot these points over the contours of gas mass.

The escape velocity from the galaxy as a function of height above the disk is shown by a red dash-dotted line.  We see that most of the stars are moving more slowly than the bulk of the stripped gas at that height.  This is because the stars are moving at the velocity of the gas when they are formed, but then are no longer accelerated by the ICM wind and begin to slow down due to the galaxy's potential.  Most of the stars with negative velocities are near the disk, but some stars out to $\sim$60 kpc have negative velocities, indicating that they are falling back onto the disk.  

If we ran the simulation for long enough (inside a large box) we would expect all of the stars below the red line to begin to eventually fall back towards the disk.  For most of these stars to be tidally stripped by the cluster potential (which we have not considered so far), the tidal radius would need to be about 60 kpc.  If our galaxy were in the Virgo cluster, which has a velocity dispersion of about 700 km s$^{-1}$, it would therefore need to be about 200 kpc from the cluster centre.  In Coma, which has a velocity dispersion of about 1000 km s$^{-1}$, the galaxy would need to be about 300 kpc from the cluster centre for tidal stripping to unbind a significant number of stars. 

\begin{figure*}
\includegraphics{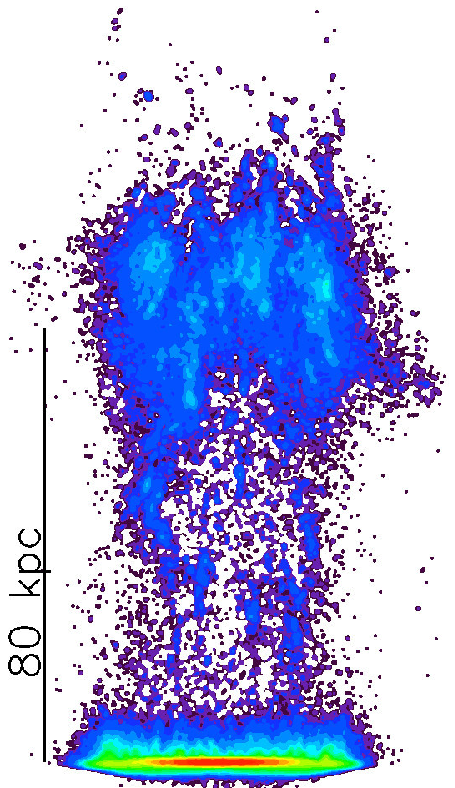}
\includegraphics{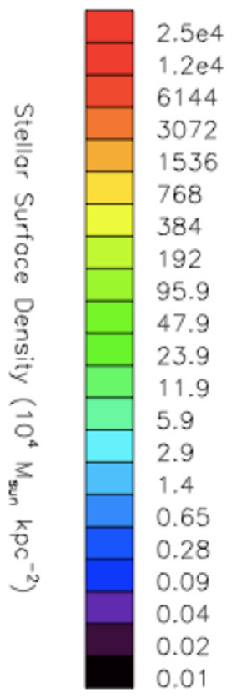}
\includegraphics{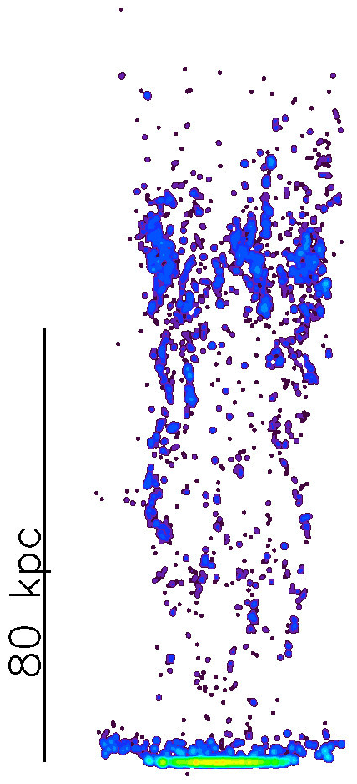}
\includegraphics{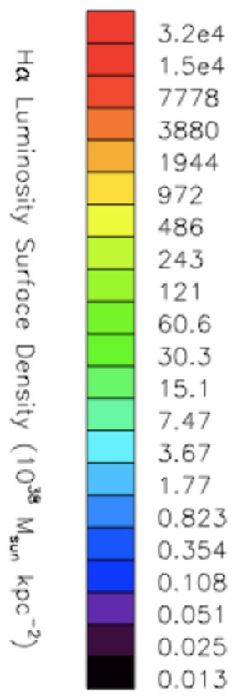}
\caption{Projections of stellar surface density on the left and the surface brightness of H$\alpha$ in HII regions on the right (computed from stars formed in the last 10 Myr, see text).  We smooth both projections using a 1 kpc gaussian.   Both old (stars formed within the stripped gas but older than 10 Myr) and new stars are well-distributed throughout the tail, reflecting the range of densities and dynamical times of stripped gas.}\label{fig-starproj}
\end{figure*}

In Figure \ref{fig-starproj} we show projections of the stellar surface density and the surface brightness of H$\alpha$ from HII regions.  First we focus on the left panel, the stellar mass surface density.  We see that there are a few clumps of $\sim$ 3 $\times$ 10$^4$ M$_\odot$ kpc$^{-2}$, which are aligned with where the recent star formation has taken place (compare to the right panel).  There is also a more diffuse component with surface densities about an order of magnitude less.  We can estimate if we should see these tails in deep images of clusters.  Each star particle is the size of a small cluster of stars that is formed at the same time using a Salpeter mass function ranging from 0.1-100 M$_\odot$.  Mengel et al. (2002) find the L$_v$/M for young star clusters to range from 0.5-2 for ages ranging from 10$^6$-10$^8$ yr.  If we assume the highest L$_v$/M, the surface brightness of the bright knots of $\sim$ 3 $\times$ 10$^4$ M$_\odot$ kpc$^{-2}$ is $\sim$29.5 mag/arcsec$^2$.  This is well within the range of V-band surface brightness of the ICL, and dimmer than the ICL observed by Mihos et al. (2005).  The dimmer, more diffuse stellar component that is clearly connected to the disk would be very difficult to distinguish from the general ICL.  

In the right panel of Figure \ref{fig-starproj} we show a projection of the surface brightness of H$\alpha$ from HII regions.  Our simulation outputs the mass and formation time of each star particle, from which it is easy to compute a SFR.  The simulation does not directly calculate an H$\alpha$ luminosity, so in order to compare with observations of HII regions (which are observationally distinguished from diffuse H$\alpha$ emission because they are small bright regions), we must assume that newly-formed (within 10 Myr) star particles produce H$\alpha$ emission from HII regions.  We use the observationally determined relation between the (recent) star formation rate and H$\alpha$ luminosity from Kennicutt (1998): SFR $(M_\odot {\rm yr}^{-1}) = 7.9 \times 10^{-42} L($H$\alpha)$ (erg s$^{-1})$.  In this calculation we are simply assuming that H$\alpha$ emission only measures the SFR within the last $\sim$ 10 Myr.  We compute the equivalent star formation rate by selecting only star particles which are younger than 10 Myr (and hence will have associated gas and bright H$\alpha$ emission).

\begin{figure}
\includegraphics{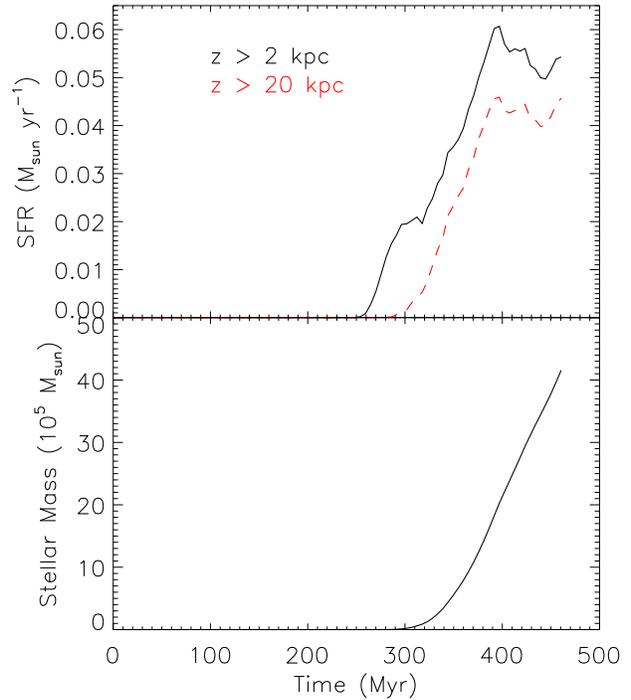}
\caption{The top panel shows the SFR in the tail gas either above 2 kpc (solid black line) or above 20 kpc (dashed red line).  There is star formation throughout the tail.  At early times the SFR rate is very low, in rough agreement with the observations of Gerhard et al. (2002) and Cortese et al. (2003; 2004).  The bottom panel plots the amount of stellar mass more than 20 kpc above the disk vs time.  Although there certainly are stars in the tail, this amount of stellar mass will not be a large fraction of the ICL even if it all escapes into the ICM.}\label{fig-icl}
\end{figure}

As in previous work (Tonnesen \& Bryan 2009; 2010), we find that ram pressure stripping cannot remove the densest clouds in the disc without breaking them apart.  Stars formed in the tail must therefore be created from the less dense gas that has cooled and condensed in the tail.  Disk gas (that is not in dense clouds) with a range of densities is stripped continuously, so there is a range of gas densities throughout the tail.  Accordingly, the time it takes for radiative cooling and compression by the ICM, and consequent star formation to occur, varies throughout the tail as well.  Figures \ref{fig-rhotdisk} and \ref{fig-rhovzdisk} show that gas with $\rho$ $<$ 10$^{-24}$ can be stripped from the disk and that some gas at these lower densities has T $<$ 10$^5$.  Because of these low temperatures, this gas will cool and condense into clouds rather than mix into the ICM (as discussed in Tonnesen \& Bryan 2010, 2011).  The exact temperature and density of the gas will determine how long this takes, which is why there is a large spread in the height of stars above the disk.  This is illustrated in the two lines in the upper panel of Figure \ref{fig-icl}, which shows the SFR in the tail above either 2 kpc (the black solid line), or 20 kpc (the red dashed line).  Star formation in the tail occurs throughout the simulation from close to the disk ($\sim$ 2 kpc) to far from the disk (well beyond 20 kpc, see also Figure \ref{fig-starproj}).  Hester et al. (2010) also find that this scenario of star-forming clouds condensing within the stripped tail agrees well with their observation of a tail from a galaxy in the Virgo cluster.

There are clearly stars in the tail, and if unbound, they can contribute to the ICL.  We evaluate this possibility with Figure \ref{fig-icl}, which shows the cumulative amount of stars formed in the tail.  We find that 250 Myr after the SFW galaxy has begun to be stripped by the ICM wind, there is about 4.2 $\times$ 10$^6$ M$_\odot$ of stellar mass more than 20 kpc above the galaxy.  From Figure \ref{fig-starpos} we know that this is an overestimate of the number of stars that will escape this galaxy.  Therefore we find it unlikely that ram pressure stripping is a large contributor to the ICL, as 4.2 $\times$ 10$^6$ M$_\odot$ is less than 1\% of the stellar mass formed in the disk.  

\section{Discussion}

\subsection{Comparison with previous work}\label{sec:simcomp}

We compare our results with the simulations in K09, as they both make predictions for star formation in the tails of ram-pressure stripped galaxies.  As we discuss in the introduction, K09 ran 12 simulations of ram pressure stripped galaxies, varying the ICM density and velocity.   Our simulation is most similar to their run 2, which has an ICM density slightly above ours (5 $\times$ 10$^{-28}$ g cm$^{-3}$ rather than 3.2 $\times$ 10$^{-28}$ g cm$^{-3}$) and a wind velocity slightly below ours (1000 km s$^{-1}$ vs 1413 km s$^{-1}$).  The amount of gas stripped is very similar -- in 250 Myr about 65\% of the original gas mass is stripped from the K09 simulation (their figure 15), while our SFW run has about 58\% of the gas mass stripped (Figure \ref{fig-gasmass}).  Given that the SFW galaxy is more massive than the K09 galaxy, this is reasonably good agreement.  K09 find that tail gas with T $>$ 10$^6$ K has a mean density of about 10$^{-24}$ g cm$^{-3}$, while our mean density is lower, at $\sim$ 2 $\times$ 10$^{-25}$ g cm$^{-3}$.  This is both because our surrounding ICM density is lower and because we are including all gas with a tracer fraction above 0.6 in this measurement, which we would expect to skew our results to lower densities.  The results regarding the stripped gas are similar between SFW and K09.  

However, our star formation results are very different from the K09 results.  In direct opposition to our results, they find that adding a ram pressure stripping wind results in more stars being formed in the simulation.  In the run most similar to ours, after 250 Myr there are nearly as many stars formed in the wake as in the disk, while in our SFW run only about 1\% of the new stars formed are in the tail.

We checked whether a difference in our threshold density for star formation could have a large effect on the star formation rate in our simulations.  To do this, we ran a comparison simulation identical to SFW in which we allowed stars to form if gas had a density above 3.85 $\times$ 10$^{-26}$ g cm$^{-3}$, a factor of 10 below our standard prescription.  The results were quite similar to our standard run.  The SFR in the disk had differences of less than 10\% at every output.  After 460 Myr, the total stellar mass in the disk with the lower density star-formation threshold was only 4\% larger than in the SFW disk, and the amount in the bulge was only 4\% lower than in the SFW bulge.  The tail results differed a bit more: about 50\% more stars were formed in the tail over the length of the simulation.  This makes sense -- decreasing the star formation threshold substantially increases the amount of gas that can form stars (see Figure \ref{fig-rhot}), but the gas has a longer dynamical time so the net change in star formation is not large.  This increase in stellar mass formed in the stripped tail is not nearly large enough to account for the difference between our results and those of K09.  Increasing our star formation efficiency would probably increase the stellar mass in the tail, but this could affect our agreement with the empirical Schmidt-Kennicutt Law.  

Our simulations are very different, so there are many possible reasons for different results, and we will list a few of the most salient differences now.  First, K09 use GADGET-2, an SPH code, and include a subgrid model that increases the gas pressure at high density.  Their galaxy is less massive than ours, with a circular velocity of 160 km s$^{-1}$.  It is also more gas-rich, with a gas fraction of 25\% of the total disk mass.  They allow radiative cooling to 10$^4$ K, and the maximum temperature at which gas can form stars is 10$^6$ K.  Their star formation prescription, like ours, is proportional to the dynamical time of the gas.  The reason for the very different predicted star formation rates in the tail is hard to pin down, but we suggest two key differences.  First, the subgrid model in the Springel \& Hernquist (2003) prescription has a very stiff equation of state in dense gas, while our dense gas typically is cold and has a low thermal pressure.  Second, inspection of the K09 results suggests that stripped gas does not mix with the ICM, resulting in a high fraction of the stripped gas ending in large, cold clumps.  We note that SPH has difficulties in resolving instabilities at interfaces (Agertz et al. 2007), resulting in undermixing and reduced stripping.  This, combined with the stiff equation of state and the high ICM pressure, lead to large star formation rates in the stripped gas.

\subsection{Comparison With Observations}\label{sec:obs}

We will first compare our results to observations of ram pressure stripped galaxies in the Virgo cluster, which have largely been identified due to their \ion{H}{I} tails (but have relatively little star formation).  We will then compare our results to observations of \ion{H}{II} regions and/or stellar tails associated with galaxies that are likely undergoing ram pressure stripping in more massive clusters.  Finally, we discuss the key physics that control the star formation rate in our simulated tails.

\subsubsection{Virgo Tails}\label{sec:virgo}

As we have discussed in the Introduction, eleven galaxies in Virgo have clear stripping signatures in \ion{H}{I} observations, but only four have been found to have star formation either from UV emission or \ion{H}{II} regions.  This begs the question of whether we should expect star formation in our stripped tail.  To answer this question we will compare our results to some of these observations.  

Let us first consider the galaxies that have star formation in their tails.  Cortese et al. (2003) found an \ion{H}{II} region 3 kpc from the disk in the stripped tail of NGC 4402.  Their measured H$\alpha$ luminosity results in a SFR (using the Kennicutt (1998) equation) of 2.3 $\times$ 10$^{-3}$ M$_\odot$ yr$^{-1}$.  Our simulated ram pressures are similar to that likely experienced by Virgo galaxies, and we have SFRs of tail gas (more than 2 kpc above the disk) ranging up to 6 $\times$ 10$^{-2}$ M$_\odot$ yr$^{-1}$ (over the 250 Myr that the galaxy is being stripped).  If we choose an early output, our SFR agrees with the observations of Cortese et al. (2003; 2004).  However, Crowl \& Kenney (2008) use stellar populations in the disk to predict that this galaxy has been stripped for nearly 200 Myr.  This means that we overpredict both the SFR and the distance to which star formation would be observed in this galaxy.  Of course, because we are not actually modeling this galaxy, our galaxy velocity is slightly higher than that expected by Crowl et al. (2005), and our galaxy is being stripped face-on rather than at an angle.  Our ICM density is very similar to that near NGC 4402 (Schindler et al. 1999).  Similarly, NGC 4330, NGC 4522, and NGC 4438 have ongoing star formation in their stripped tails (Abramson et al. 2011; Kenney \& Koopmann 1999; Boselli et al. 2005).  As in NGC 4402, the observed stars in the tails are closer to the galaxy than the more extended tails we simulate.  

Finally, we discuss IC 3418, which has not been observed in \ion{H}{I}, but has a UV and H$\alpha$ tail (Martin et al. 2005; Hester et al. 2010).  IC 3418 is likely to be in a higher-density ICM than we model by about a factor of 5.  This UV tail extends 17 kpc from the disk, and the authors calculate a lower limit for the SFR of 6 $\times$ 10$^{-3}$ M$_\odot$ yr$^{-1}$ (because they do not correct for any dust extinction).  They use the star formation truncation time to estimate that this galaxy has been stripped for 100 Myr.  After 100 Myr of stripping, our stellar tail is about 20 kpc, in good agreement with this observation.  However, our simulated SFR is still a factor of $\sim$3 above that in IC 3418.   

There are, in addition, seven galaxies that do not have any stellar light associated with their \ion{H}{I} tails.  Six of these galaxies are at or beyond 700 kpc from M87, and so may be in lower-pressure ICM regions than we simulate.  The exception is NGC 4388, which, based on the models of Vollmer \& Huchtmeier (2003), may be in a similar, or higher, density region of the ICM than we use in our simulations.  

Although we are not attempting to directly model any single galaxy, in general we have a higher SFR in our tail and a longer stellar tail than in most of these galaxies.  There are a number of reasons we find higher star formation in our stripped tail.  (i) We may be overestimating the amount of star formation in the tail due to our star formation method.  We could lower the SFR in our tail by changing our star formation criteria--while we saw in Section \ref{sec:simcomp} that lowering the star formation density threshold by an order of magnitude only changed the amount of star formation by 50\%, we could raise the threshold to the point where there was very little star formation in the tail.   (ii) We may also be overestimating the survival of star-forming clouds to large distances above the disk, a point we will discuss in more detail in Section \ref{sec:cloudsurvival}.  (iii) We may also have more star formation in our tail because we have a face-on wind that can strip more gas, or (iv) because we are modeling a higher-pressure ICM than surrounds most of the observed galaxies with \ion{H}{I} tails in Virgo.  This point will be discussed in more detail below.

\subsubsection{Stellar Tails in Massive Clusters}\label{sec:stellartails}

Turning to more massive clusters, Yoshida et al. (2008) observed ``fireballs" around a merger galaxy in the Coma cluster.  They find that these blue or H$\alpha$ emitting filaments are found on one side of the galaxy, extending up to 80 kpc from the disk.  The morphology of this tail of ``fireballs" is roughly in agreement with our simulation.  They find that the H$\alpha$ knots and filaments are farther from the disk than the blue knots and filaments, which also tends to be the case in our tail (Figure \ref{fig-starproj}), but is not as clear-cut as in the case of RB 199.  However, they only find 13 knots and filaments, while it is clear that we have many more star particles.  Further, they estimate the mass of the stars in their tail to add up to 10$^8$ M$_\odot$--at least a factor of 25 larger than the stellar mass in our stripped tail (but see below).

Recently Yagi et al. (2010) observed 14 galaxies with stellar H$\alpha$ clouds in tails in the Coma cluster.  Their images show a range of possible tails, some of which look more like our simulated tail than others.  A number of their tails show very linear trails of either young stars or \ion{H}{II} regions, which is in qualitative agreement with the right panel of Figure \ref{fig-starproj}.  Finally, Sun et al. (2007) focus on the \ion{H}{II} regions in the stripped tail of ESO 137-001.  They find 29 \ion{H}{II} regions with H$\alpha$ luminosities ranging from 10$^{38}$ to 10$^{40}$ erg s$^{-1}$.  They calculate that the total mass of all the \ion{H}{II} regions should be about 10$^7$ M$_\odot$, and using the Kennicutt (1998) relationship between H$\alpha$ luminosity and SFR estimate an instantaneous SFR of about 0.7 M$_\odot$ yr$^{-1}$.  This includes a bright \ion{H}{II} region slightly less than 2 kpc above the disk, and without this \ion{H}{II} region the SFR would be about 0.53 M$_\odot$ yr$^{-1}$.  Both are larger than what we find, by about an order of magnitude.

Why do we predict significantly less star formation in the stripped tail than in these observations?  Yoshida et al. (2008) determine an ICM density much like the one in our simulation, so there should not be an increased SFR in tail gas due to higher ICM density.  A possible explanation is that RB 199 has the disturbed morphology of a merger remnant, which may have moved large star-forming clouds farther from the centre of the galaxy to regions where they could be more easily stripped by ram pressure.

We may have less stellar mass in our tails than in ESO 137-001 and most of the galaxies observed by Yagi et al. (2010) because we simulate a lower density ICM than in those regions of the Coma and Norma clusters.  Kapferer et al. (2009) found that increasing the ICM density increased the star formation in the tail, which might cause stripped clouds to be compressed more quickly.  This agrees with observations of the lower H$\alpha$ luminosities of \ion{H}{II} regions in ram pressure stripped galaxies in the lower-density ICM of the Virgo cluster (Kenney \& Koopmann 1999; Cortese et al. 2004), as highlighted in the previous section.

In order to test this idea using our simulations, we can make a rough estimate of the star formation rate of gas in the tail in our two simulations modeled after ESO 137-001, and examined in detail in Tonnesen et al. (2011) (running these simulations including star formation at the same resolution as SFW is too computationally costly).  From equation (1) we can calculate the star formation rate simply by knowing the gas mass and density in a cell, using $t_{dyn} = (3\pi/ 32G\rho)^{1/2}$ and recalling that we have an efficiency of 0.5\%.  We first test this method by comparing the star formation rate estimated in this way to the measured rate for SFNW and SFW.  First, we consider just the disk gas, and find that, averaged over time, the estimated rate is within 1\% of the measured rate, with a variation ranging from 10\% too large, to 2\% too small.  More importantly, in the tail of the SFW run, the estimated rate is within about 10\% of the measured rate, with a similar level of scatter.

Now that we have confirmed that the errors in our estimation scheme are low in comparison to the measured rates, we can predict the star formation rate in our two simulations using similar ICM conditions to those around ESO 137-001 (Sun et al. 2006; 2010; see Tonnesen et al. 2011 for details of these runs).  Using the gas between 2 and 40 kpc above the disk in order to match the observations of Sun et al. (2006), we predict a SFR of 0.094 M$_\odot$ yr$^{-1}$ (for the T3vl run, which had somewhat lower ram and thermal pressures) and 0.32 M$_\odot$ yr$^{-1}$ (for the T3vh run, which had higher pressures).  We see that the T3vh estimate is within a factor of $\sim$2.5 of the observed Sun et al. (2007) value of 0.7 M$_\odot$ yr$^{-1}$ (or within a factor of 2 of the corrected value of 0.53 M$_\odot$ yr$^{-1}$, beyond 2 kpc above the disk), while the T3vl estimate is in poorer agreement.  As discussed in Tonnesen et al. (2011), the T3vl case also does not agree with the non-detections of \ion{H}{I} in the tail, and these star formation estimates lend more credibility to our claim that T3vh is a stronger match to the observations of ESO 137-001.  We also predict that there will be star formation in the stripped tail between 40-80 kpc with a  SFR of about 0.075 (T3vh, 0.51 for T3vl) M$_\odot$ yr$^{-1}$, which, if correct, will make it very difficult to observe.

\subsubsection{What drives the rate of star formation in the tail?}

Finally, in this section we try to determine the key physical effect that determines the star formation rate in the tail.  To do this, we compare the T3vl and SFW runs -- the velocities in the two runs are similar, and although the different ram pressures (T3vl is about a factor of 10 larger than the SFW run) lead to different mass loss rates, we can compare the two runs when there are similar amounts of gas in the tail.   When we do that, the T3vl run still produces a much larger (estimated) star formation rate, indicating it is not simply the amount of stripped gas.  Instead, we argue that, the enhanced rate is largely due to the increased thermal pressure in the ICM.  Since the cold tail gas is largely in pressure equilibrium with the ICM, increasing the ICM pressure moves that gas to higher density (see Figure~\ref{fig-rhot}), increasing the star formation rate.  Of course, the amount of stripped gas is also important, which is controlled by the ram pressure strength, but once the gas is stripped, it is largely the ICM pressure which controls the star formation rate in the wake.

\subsection{Star Formation Recipe}

As noted in Section \ref{sec:sf}, we allow stars to form in gas with densities greater than 3.85 $\times$ 10$^{-25}$ g cm$^{-3}$ and temperatures below 1.1 $\times$ 10$^4$ K.  We do not; however, require that the mass of the cell exceeds the Jeans mass, or that the cooling time be less than the dynamical time, as used in, for example, Tasker \& Bryan (2006).  If we used more strict star formation criteria, we would expect our star formation rate to decrease, although based on inspection of Figure \ref{fig-rhotdisk}, the density criteria would have to change by more than an order of magnitude before the mass of gas able to form stars would be significantly affected.  Nevertheless, we note that our more generous star formation criteria make our conclusions regarding intracluster light and the change of the bulge-to-disk ratio conservative.  In addition, we note that our results are based on a comparison between the SFR in SFNW and SFW and so should not be affected by our exact star formation recipe.  

Our star formation recipe may have a greater impact on the SFR we measure in the stripped tail of gas.  We discuss this in Section \ref{sec:simcomp}, but it is worth reiterating here.  If we used more strict star formation criteria, we may indeed see less star formation in the tail, possibly in better agreement with observations of Virgo galaxies.  While it is possible to change our calculated values of the SFR by changing our star formation recipe, our conclusion that higher ICM pressure results in more star formation is robust.

\subsection{Resolution}

\subsubsection{Star Formation}\label{sec:starres}

This work has included star formation in order to determine how ram pressure stripping affects star formation rates in the disk and tail of a stripped galaxy.  Therefore it is important to discuss how resolution may affect our results.  As we discussed above (Section \ref{sec:sf}), star formation can occur in gas with densities greater than 3.85 $\times$ 10$^{-25}$ g cm$^{-3}$ and temperatures below 1.1 $\times$ 10$^4$ K.  This corresponds to a Jeans length of about 1.9 kpc, which is resolved by 50 cells at the finest level of resolution.  These cases meet the Truelove criterion (Truelove et al. 1997), which requires a minimum of four cells per Jeans length.  However, there is some gas with much higher densities and lower temperatures in both the disk and tail (Figures \ref{fig-rhotdisk} \& \ref{fig-rhot}).  Some of this high density gas has a Jeans length of less than 10 pc, so our simulations may include artificial fragmentation that could increase our star formation rate.  Our densest gas is found in the galaxy disk, and we have found that our galaxies do lie on the Kennicutt-Schmidt relation.  Further, our newly formed stellar mass closely matches that predicted by Equation \ref{eq:sfmass}.  However, this does not prove that we do not have artificial fragmentation on small scales in high-density clouds, and so our exact measures for the SFR in the galaxy disks could be incorrect.  The comparisons between the SFNW and SFW disks should not be affected.  The tail gas has a much lower density and will largely not be affected by artificial fragmentation.  The actual values for the SFR found in the tail should be more reliable, which is most pertinent for this paper.

In galactic disks, improving the resolution allows the gas to fragment faster and into smaller clumps, leading to higher densities and larger star formation rates.    We found this effect in Tonnesen \& Bryan (2009), when we simulated disks with 20, 40 and 80 pc minimum resolution.  We found that high density gas formed faster, and the highest density reached was larger as the resolution increased, although by 40 pc, gas easily reached star forming densities. This effect was also demonstrated explicitly in Teyssier et al. 2010, who carried out merger simulations with low and high resolution, arguing that high resolution was required to correctly obtain the star formation rate.  Although their high-resolution simulation had significantly better spatial resolution (10 pc) than described here, it had significantly worse mass resolution ($4 \times 10^4 \msun$ compared to $5 \times 10^3 \msun$).  Since mass resolution is important in resolving clump formation, we conclude that our simulations probably do not significantly underestimate the star formation rate.  This is in agreement with the results of Tasker \& Bryan (2006), who carried out simulations with 25 and 50 pc, and found that although the higher resolution simulations produced small clumps, the net star formation rate was similar in the two runs.  At our current resolution (38 pc) we see individual clumps in both the disks and tails (our spatial and mass resolution is intermediate between their two resolution runs).  

\subsubsection{Cloud Survival}\label{sec:cloudsurvival}

Although we refer the reader to our previous papers for an in-depth discussion of how resolution may effect cloud survival (e.g. Tonnesen et al. 2011), it is important to note that we do not fully resolve the dense clouds in our disk or in the stripped tail.  We have found that decreasing the resolution decreases the number of dense clouds formed, and the maximum density of the clouds formed.  Also, based on the fact that a lower resolution simulation discussed in Tonnesen et al. (2011) emits less H$\alpha$ per cloud, we expect that the edges of the clouds are more diffuse with lower resolution.  The density gradient at the edge of a cloud may affect its survival, as discussed in detail in Nakamura et al. (2006) and Yirak et al. (2009), who find that a low density gradient results in slower growth of instabilities, which can retard cloud destruction.  Deep, high resolution observations in \ion{H}{I} of tails that have young stars and \ion{H}{II} regions will allow us to determine how many dense clouds form, and how long they survive, in order to better constrain our simulations.

\section{Conclusions}

We have run high-resolution galaxy simulations including radiative cooling and star formation with thermal feedback in order to understand how ram pressure stripping can influence the stellar disk and whether ram pressure stripping can contribute a large fraction of stars to the ICL.  Our main conclusions are:

1.  Including star formation and thermal feedback does not significantly affect the remaining gas disk and the stripped gas tail.  The results from our previous simulations without star formation do not need to be much modified when star formation is included.  A possible caveat to this is that star formation and feedback can affect the density distribution of very dense gas, with $\rho >$ 10$^{-23}$ g cm$^{-3}$ (see Figure \ref{fig-rhotdisk}, where the gas is puffed up and spread to lower densities), which might then make it susceptible to very strong ram pressure -- stronger than modeled here, but see Tonnesen et al. (2011).

2.  Ram pressure stripping acts to quickly reduce the total SFR of the galaxy, with a timescale of a few hundred Myr.  We find no increase in the star formation rate either due to the increase in the surrounding pressure or the shock of the ram pressure stripping wind.  The relationship between star formation rate and gas surface density (the -Schmidt relation) remains consistent whether or not a galaxy is being ram pressure stripped.

3.  Ram pressure stripping slightly increases the stellar mass of the galactic bulge relative to a galaxy forming stars in a static ICM.  However, this effect is very small and does not significantly change the bulge-to-total ratio in our model galaxy.

4. We find that star formation does occur in the tail.  This occurs not because dense (molecular) clouds are stripped wholesale, but instead because the relatively low density gas that is stripped can cool and condense into dense clouds in the turbulent wake.  We predict both the diffuse H$\alpha$ and stellar H$\alpha$ production rates in the tail.

5. Some of the stars formed in the tail are unbound from the galaxy and will become part of the intracluster light.  However, the total stellar mass added to the ICL from stripped gas is only $\sim$4.2 $\times$ 10$^6$ M$_\odot$. 

6. We compare the star formation rate we find in the tail to observations, and find that in the stripped tails of Virgo galaxies, the observed star formation rate in the tail is lower than we predict.  In more massive clusters, the star formation rate in stripped tails can be about an order of magnitude higher than our predictions.  We argue, based on these comparisons and estimations of star formation rates in previous simulations (that varied the ram pressure and thermal pressure strengths but that did not self-consistently include star formation), that the star formation rate in the tail depends both on how much gas is stripped, and also on the ICM pressure.  Higher ICM pressures create more cold, dense gas in the wakes, resulting in higher star formation rates, consistent with observational trends.

As we discuss in the text of this paper, a few of our results may be very dependent on our galaxy model and ICM parameters.  Our third conclusion, that ram pressure does not have a large effect on the B/T ratio, may depend on our disk model.  First, if our galaxy was much less massive, then we might expect the disk to be even more quickly stripped and to dim faster, while the bulge growth may remain very similar.  This could affect the bulge-to-disk ratio.  Later-type galaxies do tend to be less massive and are the very galaxies for which it is most important that the bulge mass grows in order to produce an S0 (Solanes et al. 1989).  

In addition, as we note, a more dense and/or more quickly moving wind can strip gas with higher densities.  Therefore, although in this case including star formation had very little effect on the total amount of gas stripped from the disk, at higher ram pressures we might expect more gas to be stripped from simulated galaxies that included star formation and feedback.  It would be interesting to explore star formation and feedback in models with a range of ram pressure strengths.

Our fifth conclusion, that ram pressure stripping does not add a significant amount of stellar mass to the ICL, deserves more attention.  This simulation models ICM densities and velocities of galaxies found at about the virial radius of a $\sim$4 $\times$ 10$^{14}$ M$_\odot$ cluster (Tonnesen et al. 2007).   If we estimate the amount of star formation in our simulations from Tonnesen et al. (2011), which have a higher ICM pressure modeled after the observations by Sun et al. (2006, 2007, 2010), we predict that the SFR increases by a factor of about 10.  If we assume that this means there will be 10 times as much total stellar mass added to the ICL, we still conclude that even in the centres of large clusters like Norma (ESO 137-001 is about 200 kpc from the centre of the Norma cluster), less than 1\% of the stripped gas turns into stars to be added to the ICL.  Sun et al. (2010) argued that ram pressure stripping could contribute a significant fraction of the ICL if 10\% of stripped gas forms stars, which is clearly not predicted by our simulations.  However, we did find that increased ICM pressure does increase the star formation rate of stripped gas.  Therefore, it is possible that in regions of very high ICM pressure (T $\sim$ 10$^7$ K and $\rho$ $>$ 10$^{-26}$ g cm$^{-3}$), 10\% of stripped gas could form stars.   Arnaboldi \& Gerhard (2010; references therein) write that recent studies have found that the ICL contains between 10\% and 30\% of the stellar mass in a cluster.  The stellar mass in our simulated galaxy is 10$^{11}$ M$_\odot$, and the amount of stars added to the ICL is about 4 $\times$ 10$^6$ M$_\odot$.  Even in the centre of Norma, the stellar mass added to the ICL is likely to be less than 10$^8$ M$_\odot$.  We find that ram pressure stripping will contribute a very small fraction of the total stellar mass in the ICL.

We predict that observers will continue to find that strong features in the ICL are red, indicating stars stripped from galaxies rather than star formation in the ICL (Rudick et al. 2010).  Further, we predict that ram pressure will not generally form starburst galaxies.  Finally, the bulge in our galaxies grows through gas spiraling in through the disk.  This may lead to more rotation in the bulge, a signature that could discriminate between this bulge growth method and, for example, one directly from mergers.

\vskip 1cm
We acknowledge support from NSF grants AST-0547823, AST-0908390, and AST-1008134, as well as computational resources from NASA, the NSF Teragrid, and Columbia University's Hotfoot cluster.  We thank Jacqueline van Gorkom and Jeffrey Kenney for useful discussions.  We also thank our anonymous referee for comments and suggestions that improved the quality of this paper.

\end{document}